\documentstyle[]{mn}
\newif\ifAMStwofonts
\ifoldfss
  \ifCUPmtlplainloaded \else
    \NewTextAlphabet{textbfit} {cmbxti10} {}
    \NewTextAlphabet{textbfss} {cmssbx10} {}
    \NewMathAlphabet{mathbfit} {cmbxti10} {} % for math mode
    \NewMathAlphabet{mathbfss} {cmssbx10} {} %  "   "    "
  \fi
  \ifAMStwofonts
    \ifCUPmtlplainloaded \else
      \NewSymbolFont{upmath} {eurm10}
      \NewSymbolFont{AMSa} {msam10}
      \NewMathSymbol{\upi}     {0}{upmath}{19}
      \NewMathSymbol{\umu}     {0}{upmath}{16}
      \NewMathSymbol{\upartial}{0}{upmath}{40}
      \NewMathSymbol{\leqslant}{3}{AMSa}{36}
      \NewMathSymbol{\geqslant}{3}{AMSa}{3E}

    \fi
  \fi
\fi % End of OFSS

\ifnfssone
  \newmathalphabet{\mathit}
  \addtoversion{normal}{\mathit}{cmr}{m}{it}
  \addtoversion{bold}{\mathit}{cmr}{bx}{it}
  \newmathalphabet{\mathbfit} % math mode version of \textbfit{..}
  \addtoversion{normal}{mathbfit}{cmr}{bx}{it}
  \addtoversion{bold}{\mathbfit}{cmr}{bx}{it}
  \newmathalphabet{\mathbfss} % math mode version of \textbfss{..}
  \addtoversion{normal}{\mathbfss}{cmss}{bx}{n}
  \addtoversion{bold}{\mathbfss}{cmss}{bx}{n}
  \ifAMStwofonts
    \ifCUPmtlplainloaded \else
      \UseAMStwoboldmath
      \makeatletter
      \new@mathgroup\upmath@group
      \define@mathgroup\mv@normal\upmath@group{eur}{m}{n}
      \define@mathgroup\mv@bold\upmath@group{eur}{b}{n}
      \edef\UPM{\hexnumber\upmath@group}
      \new@mathgroup\amsa@group
      \define@mathgroup\mv@normal\amsa@group{msa}{m}{n}
      \define@mathgroup\mv@bold\amsa@group{msa}{m}{n}
      \edef\AMSa{\hexnumber\amsa@group}
      \makeatother
      \mathchardef\upi="0\UPM19
      \mathchardef\umu="0\UPM16
      \mathchardef\upartial="0\UPM40
      \mathchardef\leqslant="3\AMSa36
      \mathchardef\geqslant="3\AMSa3E
    \fi
  \fi
\fi % End of NFSS release 1

\ifnfsstwo
  \DeclareMathAlphabet{\mathbfit}{OT1}{cmr}{bx}{it}
  \SetMathAlphabet\mathbfit{bold}{OT1}{cmr}{bx}{it}
  \DeclareMathAlphabet{\mathbfss}{OT1}{cmss}{bx}{n}
  \SetMathAlphabet\mathbfss{bold}{OT1}{cmss}{bx}{n}
  \ifAMStwofonts
    \ifCUPmtlplainloaded \else
      \DeclareSymbolFont{UPM}{U}{eur}{m}{n}
      \SetSymbolFont{UPM}{bold}{U}{eur}{b}{n}
      \DeclareSymbolFont{AMSa}{U}{msa}{m}{n}
      \DeclareMathSymbol{\upi}{0}{UPM}{"19}
      \DeclareMathSymbol{\umu}{0}{UPM}{"16}
      \DeclareMathSymbol{\upartial}{0}{UPM}{"40}
      \DeclareMathSymbol{\leqslant}{3}{AMSa}{"36}
      \DeclareMathSymbol{\geqslant}{3}{AMSa}{"3E}
    \fi
  \fi
\fi % End of NFSS release 2

\ifCUPmtlplainloaded \else
  \ifAMStwofonts \else % If no AMS fonts
    \def\upi{\pi}
    \def\umu{\mu}
    \def\upartial{\partial}
  \fi
\fi

\title[A measurement of the transverse velocity of Q2237+0305]
  { A measurement of the transverse velocity of Q2237+0305 }
\author[J. S. B. Wyithe, R. L. Webster \& E. L. Turner]
  {J.~S.~B.~Wyithe $^1$, 
  R.~L.~Webster $^1$,
  E. L.~Turner $^2$\\
  $^1$ School of Physics, University of Melbourne, Parkville, Vic, 3052, 
Australia\\
  $^2$ Princeton University Observatory, Peyton Hall, Princeton, NJ 08544 USA\\ 
 Email: swyithe@physics.unimelb.edu.au, rwebster@physics.unimelb.edu.au, elt@astro.princeton.edu }
\date{Accepted  Received }
\pagerange{\pageref{firstpage}--\pageref{lastpage}}
\pubyear{1999}

\def\LaTeX{L\kern-.36em\raise.3ex\hbox{a}\kern-.15em
    T\kern-.1667em\lower.7ex\hbox{E}\kern-.125emX}

\begin{document}

\label{firstpage}

\maketitle

\begin{abstract}

Determination of microlensing parameters in the gravitationally lensed quasar Q2237+0305 from the statistics of high magnification events will require monitoring for more than 100 years (Wambsganss, Paczynski \& Schneider 1990). However we show that the effective transverse velocity of the lensing galaxy can be determined on a more realistic time-scale through consideration of the distribution of light-curve derivatives. The 10 years of existing monitoring data for Q2237+0305 are analysed. These data display strong evidence for microlensing that is not associated with a high magnification event. An upper limit of $v_{t}< 500\,km\,sec^{-1}$ is obtained for the galactic transverse velocity which is smaller than previously assumed values. The analysis suggests that the observed microlensing variation may be predominantly due to stellar proper motions. The statistical significance of the results obtained from our method will be increased by the addition of data points from current and future monitoring campaigns. However reduced photometric errors will be more valuable than an increased sampling rate.

\end{abstract}


\section{Introduction}

Q2237+0305 is a quasar at redshift $z$=1.695 that is gravitationally lensed by a foreground spiral galaxy ($z$=0.0394). Due to the unusually close alignment, the quasar is separated into 4 resolved images having separations of $\sim1''$. These images are seen through the bulge of the lensing galaxy, and so Q2237+0305 is an excellent candidate for the observation of gravitational microlensing.
Since the discovery of microlensing in Q2237+0305 (Irwin et al. 1989, Corrigan et al. 1991) various numerical techniques have been used  to model the observed continuum flux variations (eg. Wambsganss, Paczynski \& Katz 1989; Witt, Kayser \& Refsdal 1993). The models demonstrate that it will be possible to use the statistics of high magnification events (HMEs) from monitoring over a large period of time ($>100$ years) (Wambsganss, Paczynski \& Schneider 1990) to determine properties of the lens such as the the stellar mass function and the percentage of mass in compact objects. 

The most important unknown parameter in microlensing models is the galactic transverse velocity. In the case of Q2237+0305 most previous analyses have assumed a value of $v_{t}=600\,km\,sec^{-1}$. The uncertainty in the value of transverse velocity makes the analysis of monitoring data difficult because statistics of HMEs are proportional to its value. In addition, since the characteristic length or Einstein radius is proportional to the square root of the mass, attempts to obtain information on the typical mass of microlenses are dependent on the square of $v_{t}$. Witt \& Mao (1994) have found a height-gradient correlation during model HMEs. This correlation is dependent on the source size assumed, however they have interpreted the observed candidate HMEs in light of this correlation and find a transverse velocity that is lower than $600\,km\,sec^{-1}$. The rate of rise or fall during the initial or final stages of the HME is dependent on the angle between the trajectory and the relevant caustic, the source size and profile, as well as the transverse velocity. At least part of this degeneracy will need to be broken in order for a successful analysis of any HME to be made.

The contribution to microlensing of proper motions of microlenses has not been included in the aforementioned analyses. This neglect is due to the computational difficulties involved in the calculation of light curves from models which include stellar proper motions. The assumption of a static lens configuration requires an effective galactic transverse velocity that is an order of magnitude or so higher than the typical random motion. This is not a good assumption however if the unknown transverse velocity is of a similar magnitude or smaller than the stellar velocity dispersion.

The effect of proper motions on microlensing has been discussed by several authors (Schramm et al. 1992; Kundic \& Wambsganss 1993; Kundic, Witt \& Chang 1993; Wambsganss \& Kundic 1995). Kundic \& Wambsganss 1993 and Kundic, Witt \& Chang 1993 have discussed the manner in which stellar proper motions increase the frequency of HMEs. They agree quantitatively on the manifestation of this effect in the case of image A of Q2237+0305, but not on how it is manifested in general. Wyithe, Webster \& Turner~(1999) considered the effect of proper motions in terms of the distribution of resulting light curve derivatives. An analysis of the expected contribution of proper motions to microlensing in Q2237+0305 is presented as part of the current work, and is based on formalism developed in that paper.   

In addition to the difficulties presented by the relatively rare occurrence of HMEs, the analysis of current monitoring data is hampered by the fact that the sparse sampling rate may have lead to HMEs being missed altogether. However, comparatively speaking the monitoring data provides a good record of the longer duration low level fluctuations which do not involve new critical images or (very rarely) a cusp. This shift in focus vastly increases the number of data points that can be analysed to determine the physical properties of the system.

In this paper we investigate the distribution of the rate of change of magnification in the light-curve, particularly in regions of the light-curve where the size of the fluctuation is relatively small compared to that during an HME. The advantages of such an approach are numerous. Firstly, far from a caustic where the rates of fluctuation are small, the light-curve is approximately independent of source size and intensity profile. Secondly, a given section of light-curve has a typical magnification that is dependent on the local region of the magnification pattern. Due to the clustered nature of the caustic network in the presence of a shear (as for Q2237+0305), the average magnification in a region of a light-curve can differ from the theoretical mean ($\mu_{av}=1/|(1-\kappa)^{2}-\gamma^{2}|$) by a significant amount for a period of up to a few decades. A plot of the derivative is not expected to exhibit clustering of this type as it is approximately independent of the number of slowly varying pairs of semi-critical images associated with a source having crossed caustics.

This paper is presented in seven parts. Section \ref{tran_velocity} shows how the transverse velocity can be estimated from the distribution of the rates of change of the difference between the image magnitudes. In sections \ref{sect_obs} and \ref{sect_phys}, the effects of observational and physical parameters on the measurement of transverse velocity are discussed. Section \ref{prop_motion} describes the effect of stellar proper motions on the derivative distribution and in section \ref{application} the theory is applied to the monitoring data of Irwin et al. (1989), Corrigan et al. (1991) and $\O$stensen et al. (1995).
The results obtained are presented in section \ref{results}.

\section{Measuring The Effective Transverse Velocity}
\label{tran_velocity}

The observed microlensing rate is produced by the combination of microlens proper motions and a galactic transverse velocity. We define the effective transverse velocity as being that which in combination with a static microlensing model, produces a microlensing rate equal to that of the observed light curve. The effective transverse velocity is therefore larger than the physical transverse velocity. This section describes the method used to estimate the effective transverse velocity from monitoring data. 

\subsection{Microlensing models}

 Throughout the paper, standard notation for gravitational lensing is used. The Einstein radius of a 1$M_{\odot} $ star in the source plane is denoted by $\eta_{o} $. The normalised shear due to external mass is denoted by $\gamma $, and the convergence or optical depth by $\kappa $. The normalised lens equation for a field of point masses with an applied shear in terms of these quantities is
\begin{equation}
\vec{y}= \left( \begin{array}{cc}
         1-\gamma & 0 \\
        0 & 1+\gamma 
            \end{array} \right)\vec{x} -\kappa_{c}\vec{x} + \sum_{N_{stars}}m_{i}\frac{(
\vec{x}_{i}-\vec{x})}{|\vec{x}_{i}-\vec{x}|^{2}}
\end{equation}
Here $\vec{x}$ and $\vec{y}$ are the normalised image and source position respectively, and the $\vec{x_{i}}$ are the normalised positions of the point masses. $\kappa_{*}$ and $\kappa_{c}$ are the optical depth in stars and smoothly distributed matter respectively. Where required, a cosmology having $\Omega=1$ with $H_{0}=75\,km\,sec^{-1}$ is assumed.

 To construct light-curves that result from microlensing of a point source we use the contouring method (Lewis, Miralda-Escude, Richardson \& Wambsganss 1993; Witt 1993). Where a light curve is required for an extended source, we use the 2 dimensional extension of the contouring method (Wyithe \& Webster 1999) or an equivalent 1-d approximation (eg. Witt \& Mao 1994). The microlensing models consist of a homogeneous disc of point masses. Katz, Balbus \& Paczynski (1986) obtained an expression describing the region of the lens plane in which image solutions need to be found to ensure that $\sim99\%$ of the total macro-image flux is recovered. This can be termed the flux collecting region. The union of flux collecting regions that correspond to each point on the source line is termed the shooting region. The dimensions of the shooting region are determined in the standard manner (eg Lewis \& Irwin 1995; Wyithe \& Webster 1999), and the disc of point masses chosen to have a radius that is 1.2$\times$ that required to cover this region. The numbers of stars used in the microlensing models under the different assumptions for the mass distribution used in this paper are given in table \ref{tab_field}.

\subsection{The Distribution of Derivatives}
\label{distribution_of_derivatives} 

\begin{figure}
\vspace*{67mm}
\includegraphics{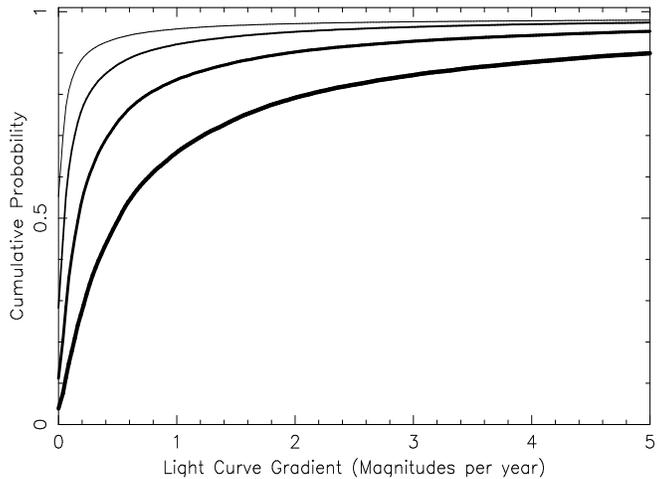}
\caption{\label{cumul_hist}The cumulative histograms calculated using the transverse speeds 100, 300, 900, and 2700$km\,sec^{-1}$ (Thinner lines denote smaller velocities). The point masses are distributed with a Salpeter mass function $p(m)dm\propto m^{-2.35}dm$ and $0.1M_{\odot}<m<1.0M_{\odot}$. The light curves were computed for a point source. The optical depth was $\kappa=0.36$, and the shear $\gamma =0.4$. }
\end{figure}

The basic tool of our analysis is the distribution of light curve derivatives in a given simulation, collection of simulations or set of observational monitoring data. The transverse velocity is a scaling factor in the light curve derivative.  Figure \ref{cumul_hist} shows the cumulative distribution functions of light curve derivatives for models of image A of Q2237+0305 that have a positive shear ($\kappa=0.36,\gamma=+0.4$) and transverse velocities ranging between $100\,km\,sec^{-1}$ and $2700\,km\,sec^{-1}$. The cumulative histogram has a smaller value if the transverse velocity is higher because at higher transverse velocities a larger proportion of points have larger derivatives. Thus by considering the shape of the cumulative distribution of light curve derivatives, information on the transverse velocity can be obtained.

\subsection{General Considerations}

As we show, treating the transverse velocity as a variable allows us to obtain limits on its value. When analysing data from a gravitationally lensed quasar such as Q2237+0305, we are interested in the independent fluctuations in the individual images. It is advantageous to look at the rates of change of the independent differences between the magnitudes of the images. There are two reasons for this. Firstly any intrinsic fluctuation that is present in the source quasar will be removed by this procedure. We note however, that in general, the time delay must be taken into account. For Q2237+0305 the delays are less than one day (Schneider et al. 1988) and so are much shorter than the
intrinsic source variability timescale. The second advantage is that when constructing a light curve from observations, the error in the photometry comprises two components (Irwin et al. 1989). The first of these is the random error due to profile fitting for the brightness at the position of each quasar image. The second error is approximately systematic across all images and is due to errors produced in the procedure of subtracting a suitably scaled model galaxy from the image. If the magnitudes in two different observations of a single image are being compared, then the systematic errors need to be taken into account. However the systematic component of the error is unimportant when comparing the difference between the magnitude of two different images at two different times. In the case of Q2237+0305 the histogram is composed of fluctuations in six differences between the image magnitudes ($A-B,A-C,A-D,B-C,B-D,C-D$). While these 6 differences still only contain 4 degrees of freedom, all 6 are needed so that the level of fluctuation measured is not biased towards any one image.

It is assumed that the macro models (eg. Schmidt, Webster \& Lewis 1998) for the lensing galaxy in the Q2237+0305 system correctly describe the shear and optical depth parameters at the position of each image. The microlensing model of Q2237+0305 includes the following unknowns as input parameters:

\noindent $i$) Transverse Velocity- This is thought to be $\sim600\,km\,sec^{-1}$, a quantity obtained through probabilistic arguments in combination with general cluster dynamics (eg. Mould et al. 1993). Witt \& Mao (1994) find a lower value, however their measurement is dependent on source size.

\noindent $ii$) Trajectory direction- The shear term at each image breaks the circular symmetry so that the direction of the transverse component of the galactic velocity will influence microlensing statistics. This direction is completely unknown.

\noindent $iii)$ Source Size- This is thought to be $<2\times 10^{15}\,cm$ (eg. Wambsganss, Paczynski \& Schneider 1990; Rauch \& Blandford 1991), however this measurement is based on one poorly sampled HME. The microlensing value is corroborated by the length scale, $ct\sim10^{14}t_{hrs}\,cm$, associated with X-ray and optical changes in the continuum emission. A source size of $\sim10^{15}cm$ is also consistent with the typical scalesize of a continuum emitting accretion disc about a super-massive black hole (Rees 1984).

\noindent $iv)$ Source Intensity Profile- This is dependent on the model assumed for the continuum source. It is only an important factor during a HME (Wyithe \& Webster 1999).

\noindent $v)$ Microlens Properties- The mass limits and mass function of the stars and compact objects, as well as the size of the smooth matter component of the optical depth in the lensing galactic bulge must be assumed based on information about these quantities in our own galaxy.

\subsection{The Kolmogorov$-$Smirnov Statistic}
\label{ks}
The comparison between monitoring data of Q2237+0305 and ensembles of simulations is made through indirect use of the Kolmogorov-Smirnov (KS) statistic. For two cumulative distributions $F$ and $G$, the KS statistic is defined:
\begin{equation}
D\equiv max(|F-G|).
\label{ks_stat}
\end{equation}
This statistic is equivalent to the probability of the null hypothesis that $F$ and $G$ are different. This equivalence relies on the data points used to build the distributions  $F$ and $G$ being independent. Unfortunately when the histograms are comprised of simulated or observed microlensing data, the component points are not independent because of the common region of caustic network from which they are drawn. The KS statistic is useful however for comparing the data to a range of simulations.

For a given transverse velocity, the time averaged distribution is produced from a large collection of 10 year sampled simulations. The KS statistic is then found between each of the simulated histograms and the time averaged histogram. This produces a cumulative distribution of KS statistics $P(D_{sim}(v_{t}))$ for that transverse velocity $v_{t}$ which describes the spread in simulations due to the combination of caustic clustering and small monitoring length. The KS statistic can then be found between the time averaged histogram and the observational histogram ($D_{obs}(v_{t})$). The suitability of the transverse velocity for describing the data set is then discussed in terms of the relationship of $D_{obs}(v_{t})$ to the simulated distribution $P(D_{sim}(v_{t}))$. At a transverse velocity $v_{t}$, $P(D_{obs}(v_{t}))$ finds the fraction of simulations that are more consistent with the mean than the observations .

To compute the upper limit of the transverse velocity it is natural to modify the KS statistic to find whether the distribution has many more high derivatives than the mean. Such a measure is:
\begin{equation}
D_{U}\equiv max(F-G),
\end{equation}
where $F$ is the time averaged histogram. Conversely, a lower limit is obtained through the statistic
\begin{equation}
D_{L}\equiv max(G-F).
\end{equation}
The normal KS statistic (Eqn \ref{ks_stat}) can be used to find the most likely transverse velocity, though this approach is much more susceptible to accepting a false hypothesis than to than rejecting a true one, and so the expected value must be treated with more caution than the limits.

\section{The Effect of Observational Parameters}
\label{sect_obs}
\subsection{The Effect of Sampling Rates and Monitoring Periods}

\begin{figure}
\vspace*{60mm}
\includegraphics{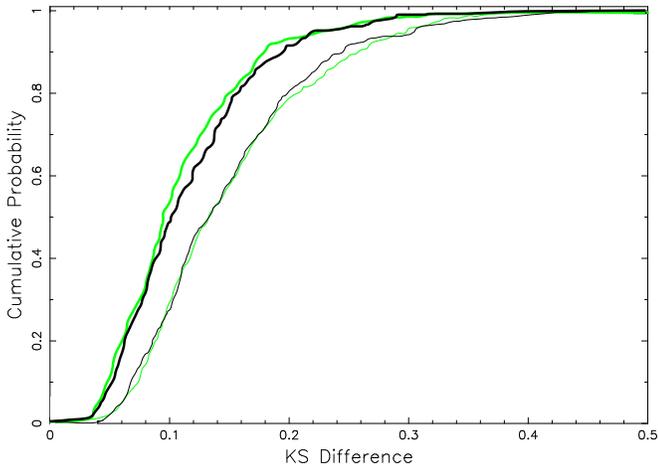}
\caption{\label{sample_KS_hist}The cumulative histograms of KS differences between time averaged and 500 sampled simulations for different sampling rates and monitoring periods. The thin lines correspond to a 10 year monitoring period, and the thick line to a 20 year monitoring period. The light lines correspond to a sampling rate of 1 point per 15 days, and the dark lines to 1 point per 30 days. Both sampling rates are for 6 months per year. The models were calculated from sampled difference light curves of models of Q2237+0305 for a point source which contained no simulated observational errors. The transverse velocity was $400\,km\,sec^{-1}$. The point masses are distributed with a Salpeter mass function $p(m)dm\propto m^{-2.35}dm$ and $0.1M_{\odot}<m<1.0M_{\odot}$. The optical depth and shear values were those of Q2237+0305, with $\gamma_{A}, \gamma_{B}>0$ and $\gamma_{C}, \gamma_{D}<0$.}
\end{figure}

When a histogram is produced from a sampled light curve, its shape is dependent on the sampling rate. In the case of a sparsely sampled curve the shape is also dependent on the type of numerical derivative computed.  A two point derivative will optimally sample noise in the data, while a higher order calculation will smear out real fluctuations due to the low sampling rates. A simulated observational error is built into the model. The greatest difference between histograms calculated for noise only and those calculated for a light-curve plus noise is obtained using the three point derivative. We therefore choose to use a three point derivative for the calculating the derivative histograms of sampled light curves.

Figure \ref{sample_KS_hist} displays examples of the distributions $P(D_{simul})$ of KS values between the simulations and time averaged distribution. The figure shows histograms corresponding to sampling rates of 1 point per 15 and 30 days, as well as monitoring periods of 10 and 20 years. 
The size of the determined range of transverse velocities is equivalent to the fraction of large KS differences found in the simulations. We find that the different sampling rates considered have very little effect on the size of the determined range of transverse velocities.
 In contrast, the length of the monitoring period has a significant effect on how good a representation a given simulation is of the time average. The longer monitoring period has a higher fraction of simulations with smaller KS differences, which indicates that at the low transverse velocities being considered here, the sampled part of the light curve has approximately the same length or is shorter than a typical cluster of caustics. This is expected since in the limit of large monitoring periods, the distribution of KS differences is a step function at zero. For the present problem it is therefore the length of the monitoring period rather than the time between observations that is more important in obtaining an accurate statistical measure.

\subsection{The Effect of Observational error}

\begin{figure}
\vspace*{60mm}
\includegraphics{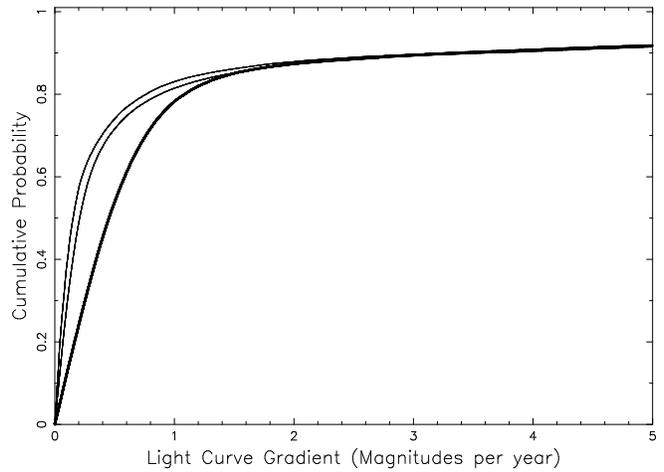}
\caption{\label{error_hist} The cumulative histograms of derivatives calculated from sampled difference light curves of models of Q2237+0305 for a point source which contain simulated observational errors. The thicker lines represent the larger errors. Three $1\sigma$ levels were used, $\Delta M = \pm 0, \pm 0.01, \pm 0.04$. The sampling rate was 1 point per 15 days for 6 months per year, and the transverse velocity was $400\,km\,sec^{-1}$. The microlensing model was the same as that described for Figure \ref{sample_KS_hist}.}
\end{figure}

Figure \ref{error_hist} displays the effect on the derivative histogram of the inclusion of observational error into the models. Observational error is simulated as an additional random component in the model light curve that is distributed according to a Gaussian. From figure \ref{error_hist} we see that the inclusion of errors has the effect of lowering the proportion of small derivatives in the light curves. This effect is the manifestation of the observational noise, and is due to the measurement uncertainties introducing independent artificial fluctuation into the flat parts of the light curves. In cases of a large observational error, the dearth of low derivatives result in a lower estimate of transverse velocity. Where the errors are extremely large, the microlensing signal will therefore be lost in the noise. This would be indicated by a measure of transverse velocity which is not significantly above zero.
 This suggests that a reasonable understanding of the size and nature of observational errors is required for a successful analysis to be made. Indeed, for a determination of transverse velocity, it is more important to have smaller errors than higher temporal resolution.
The inclusion of errors has a negligible effect on the size of the measured range of transverse velocity.

\section{The Effect of Varying Source Size}

\label{sect_phys}
The measurement of transverse velocity that we make is systematically affected by the assumptions made for the model parameters of source size, trajectory direction, and microlens mass function. The effect of the unknown source size on our measurement is discussed below. The effects of the trajectory direction, and microlens mass function are deferred to section \ref{results} where they are discussed in the context of measurements made from the monitoring data on Q2237+0305. 

\label{source_sixe}
\begin{figure}
\vspace*{60mm}
\includegraphics{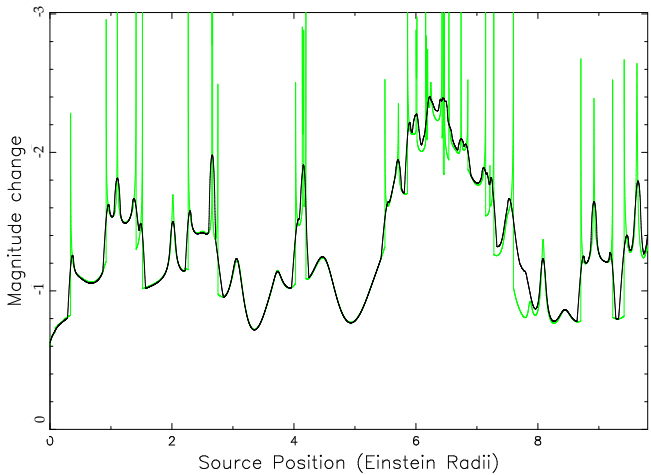}
\caption{\label{point_exten_lcurve}The microlensed light curves for a point source (light line) and $0.1\eta_{o}$ source. The extended source light curve was computed using the 2D contouring method, and has a limb darkened profile. The point masses are distributed with a Salpeter mass function $p(m)dm\propto m^{-2.35}dm$ and $0.1M_{\odot}<m<1.0M_{\odot}$. The light curves were computed for a point source. The optical depth was $\kappa=0.36$, and the shear $\gamma =+0.4$. }
\end{figure}

The detail of a light curve, particularly during HMEs is dependent on both the size and intensity profile of the source. Figure \ref{point_exten_lcurve} shows model light curves for image A of Q2237+0305 ($\gamma_{A}=+0.4$) in both the cases of a point source and a $0.1\eta_{o}$ diameter source with a limb darkened profile. It is clear from this diagram that a point source is subject to much larger fluctuations during HMEs than a larger source. The smaller source size however also means that the source spends less time in contact with a caustic so that HME scale derivatives occur less often. This results in a relative excess of the largest light curve derivatives in the case of the point and smaller sources, which is offset by an increase in the number of small light curve derivatives.

The light curve of an extended source can be produced through an appropriately weighted integration along the point source light curve (eg. Witt \& Mao 1994). Such an approximation obtains event amplitudes which are close to their correct values, although the details of the events are not accurate. The approximation assumes that the caustic is straight and perpendicular to the source trajectory. In addition it assumes that their are no other caustics within a source radius of the source line on either side. Neither of these assumptions will be true in general. We use the approximation in the current work as we are not concerned with the detail of HMEs due to the low sampling rate.

When a sampling rate is applied to the light curve, a smaller source may experience fewer large fluctuations. Depending on the set of observations, a higher transverse velocity may therefore be required to produce a given level of fluctuation and so a correspondingly smaller transverse velocity measured where a larger source size is assumed. This is in contradiction to the theoretical histogram.

\section{Microlensing Due to Stellar Proper Motions} 

\label{prop_motion}

The above models include stellar positions that remain constant during the period of the simulation. These models produce unrealistic simulations because the measured line-of-sight stellar velocity dispersion of the bulge is $\sim215\,km\,sec^{-1}$ (Foltz, Hewitt, Webster \& Lewis 1992), and models of the lensing galaxy suggest values of $\sim165km\,sec^{-1}$ (Schmidtt, Webster \& Lewis 1998). These can be compared with the expected transverse velocity of $\sim600km\,sec^{-1}$. In the remainder of this work we make two assumptions about the stellar velocity dispersion. Firstly we assume that the stellar velocity dispersion is isotropic, and secondly that a line of sight velocity dispersion of $\sim165\,km\,sec^{-1}$ is representative of its value at each of the 4 image positions.

 Theoretically the histogram of light curve derivatives provides a natural way to take account of the effect of stellar proper motions in the microlensing model. The approach is convenient because rather than having to compute many light-curves, each with a slightly evolved starfield (eg. Kundic \& Wambsganss 1993 and Wambsganss \& Kundic 1995), we only need to compute the light-curve derivative at each point. This is because for the determination of the average behaviour, the derivatives do not need to be sequential along a computed light curve. The formalism for this process is developed in Wyithe, Webster \& Turner (1999), but is described briefly below. 

Analytical expressions were obtained for the rate of change of the amplification $\frac{d\,\mu_{i}}{d\,t}$ of a given image. Through the contouring method images are found for each source point along a source trajectory. The derivatives $\frac{d\,\mu_{i}}{d\,t}$ of each image of the source are then added together to give the change in magnification:
\begin{equation}
\frac{d\,\mu_{p}}{d\,t}=\sum_{i=0}^{N_{*}}sign(\mu_{i})\frac{d\,\mu_{i}}{d\,t}
\end{equation}  
These derivatives facilitate the construction of the cumulative distribution of light curve derivatives analogous to those in figure \ref{cumul_hist}.
 As the same caustics and source are involved, we expect the cumulative distributions resulting from the two classes of motion to be similar up to a constant scaling factor in the derivative.
We define an equivalent transverse velocity to be that in a static model which produces a cumulative distribution having the smallest possible KS difference between itself and a distribution composed of derivatives resulting from a point mass velocity dispersion.

Table \ref{prop_mot_dir} shows the equivalent transverse velocity computed from the derivatives of the set of 6 difference light curves computed in the case of Q2237+0305. The histograms are very similar in form with minimised KS differences of $\sim10^{-2}$. The errors quoted (and similar errors which follow) were calculated as the standard deviation between values obtained from three separate sets of simulations. It is important to note the large contribution being made to the microlensing rate by the velocity dispersion, which is larger than that due to a transverse velocity of equal magnitude.

\begin{figure}
\vspace*{65mm}
\includegraphics{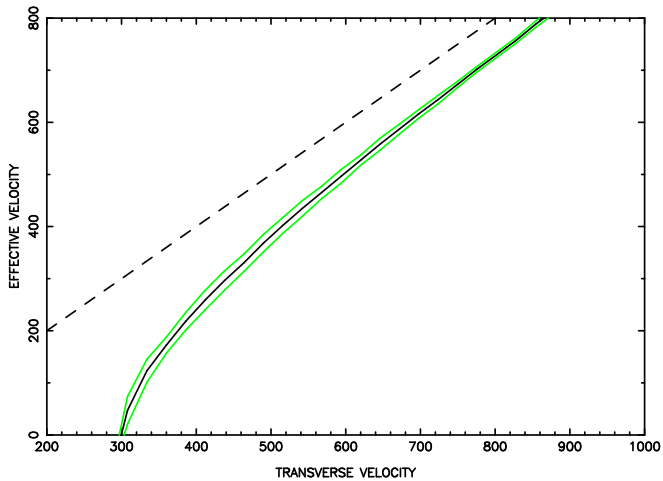}
\caption{\label{vel_convert}The relationship between the equivalent transverse velocity and the true transverse velocity (dark line) calculated from the difference light curves of models of Q2237+0305 for a point source. The light lines are the 1 standard deviation uncertainty computed from 3 simulations, and the dotted line shows the line of equality.  The point masses are distributed with a Salpeter mass function $p(m)dm\propto m^{-2.35}dm$ and $0.1M_{\odot}<m<1.0M_{\odot}$. The optical depth and shear values were those of Q2237+0305, with $\gamma_{A}, \gamma_{B}>0$ and $\gamma_{C}, \gamma_{D}<0$.}
\end{figure}

\begin{table}
\begin{center}
\caption{\label{prop_mot_dir}Table showing the relationship between the histograms of derivatives for light curves of point sources where the flux variation is due to proper motion and that where it is due to transverse velocity with a static lens. The details are described in the text}
\begin{tabular}{|c|c|c|c|}
\hline
Shear &  & Equivalent       & Equiv. Vel.   \\
      &  & Velocity         & KS difference \\ \hline\hline 
  $\gamma_{A},\gamma_{B}>0,\:\:\:   \gamma_{C},\gamma_{D}<0$   &  & 300 $\pm$ 10       &  .010         \\
  $\gamma_{A},\gamma_{B}<0,\:\:\:   \gamma_{C},\gamma_{D}>0$   &  & 280 $\pm$ 10       &  .019         \\ \hline
\end{tabular}
\end{center}
\end{table}
 
 The derivative at each point on a microlensed light curve is the result of fluctuations due to both the transverse velocity and the changes in the magnification pattern with time that result from the changing stellar positions. The derivative is additive, and this allows the theoretical histogram to be computed for the combination of the two effects by adding the derivative due to transverse velocity and that due to stellar proper motions at each point.

An analogous process to that already described allows us to find the effective transverse velocity in a static model that produces a microlensing rate equivalent to a model where proper motions and a transverse velocity are considered in combination. Figure \ref{vel_convert} shows the relationship between the effective transverse velocity and the galactic transverse velocity in a model of the 6 difference light curves of Q2237+0305. The point masses were given a 1-d velocity dispersion of $165\,km\,sec^{-1}$. In this diagram the dark line represents the mean and the light lines the $\pm$ 1 standard deviation values. The application of plots such as that shown in figure \ref{vel_convert} is to subtract off the contribution of proper motions from measurements of the effective transverse velocity that are made. This process relies on the fact that the distribution of fluctuations is independent of whether they are due to transverse velocity only, or a transverse motion in combination with stellar proper motion. In practice, account also needs to be taken of the sampling rate. However the number of caustics per unit area of the source plane is unchanged through the inclusion of proper motion. The same is true of the typical amplitude of fluctuations. Given that the distribution of the derivatives is also of the same form, we expect the sampling rate to have approximately the same effect at a given level of fluctuation in the two classes of model. The technique of conversion between an effective and a physical transverse velocity makes it practical to explore a range of microlensing models that include the stellar proper motion and compare these to observations.

\section{Application to Q2237+0305}
\label{application}

\subsection{The Monitoring Data}

\begin{figure*}
\vspace*{65mm}
\includegraphics{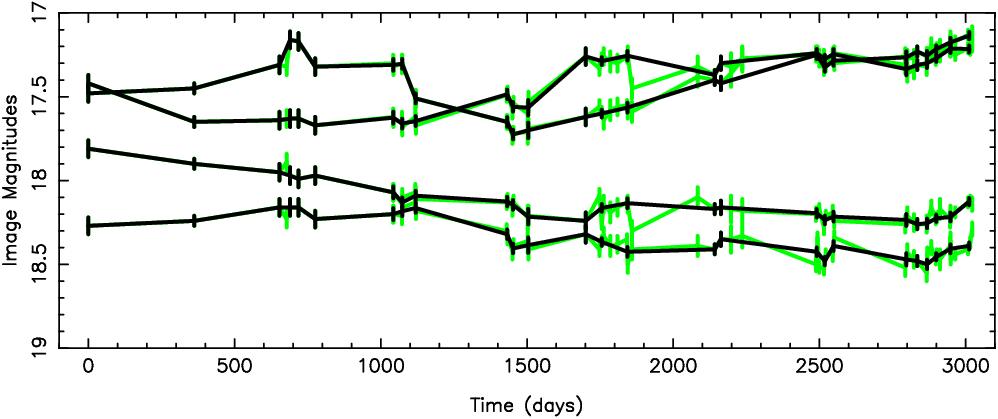}
\caption{\label{data}The observed light curve of Q2237+0305. The thin line shows the entire data set, while the thick line shows the subset of this data which has been used in this analysis. The quoted observational errors are plotted at each point.}
\vspace*{55mm}
\includegraphics{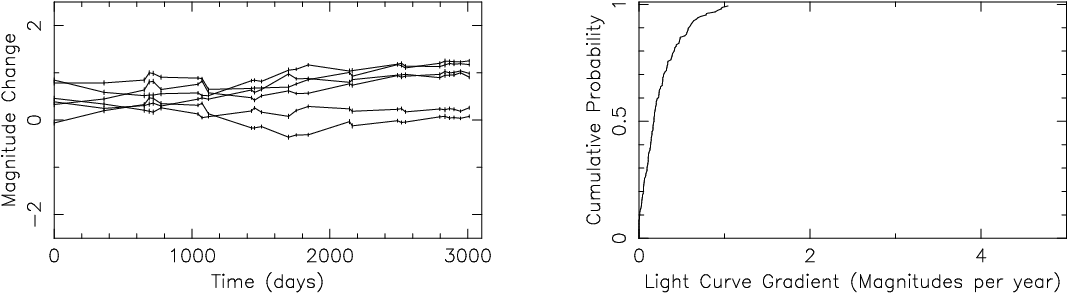}
\caption{\label{observations}Left: The six differences between images of the observed light curve of Q2237+0305. Right: The resulting cumulative histogram. The quoted observational errors are plotted at each point.}
\end{figure*}

The gravitational lens Q2237+0305 has been monitored by several groups. Kent \& Falco (1988) and Schneider et al. (1988) imaged the 4 separate quasar images and produced models which accurately reproduced their positions. Later monitoring programs were undertaken by the collaborations Irwin et al. (1989), Corrigan et al. (1991) and $\O$stensen et al.(1996). By combining this data, a light curve is produced for each of the four images. The most complete light curves are in R-band, and it is these curves that are analysed here. There are a total of 61 published data points for each image taken between 1985, and 1996. In an effort to minimise the noise in the data, any points in the sample which were taken within one week of each other were averaged and their quoted errors added in quadrature. Following this procedure, any points having an associated error above $\Delta m=0.05$ magnitudes were removed from the sample since data points with large errors substantially degrade the measurement of transverse velocity through introduction of noise into the low derivative regime. There were also two data points that were discarded since they displayed correlated flux variation in two images. This leaves 26 points. Figure \ref{data} displays the complete data set (thin line) as well as the set following modification with the above procedure (thick line). 

 Irwin et al. (1989) estimated the differential random errors in their data to be $\Delta m = \pm 0.01-0.02$. Accordingly, for this analysis we take the random error in the measured magnitude of each image to be $\Delta\,m=\pm0.02$ magnitudes. $\O$stensen et al. (1996) note that the actual statistical quality will be better for the stronger A and B components. In addition, $\O$stensen et al. (1996) find that the simultaneous increase in brightness of all four images towards the end of the monitoring period has a standard deviation of $\Delta m=0.02$ indicating that the estimates of error were indeed realistic. The errors have been assigned to the model light curves according to a Gaussian distribution. The size of the error built into the simulation has an effect on any comparison that is made with the data, so we have considered two cases. The assigned random error is taken firstly as the $2\sigma$ level in images A and B, and the $1\sigma$ level in images C and D, and secondly as the $1\sigma$ level in images A and B, and the $\frac{1}{2}\sigma$ level in images C and D. These cases are expected to bracket the uncertainty in the observed image magnitudes.      

From the manipulated data we calculate the variations in the 6 image magnitude differences (A-B, A-C, A-D, B-C, B-D, C-D). Figure \ref{observations} displays the resulting difference light curves. Microlensing in this figure is conspicuous as variation that is present in three of the six curves.  Figure \ref{observations} also shows the resulting cumulative histogram of derivatives. We consider the range of effective transverse velocities obtained through comparison of this monitoring data with various microlensing models.

\begin{figure}
\vspace*{60mm}
\includegraphics{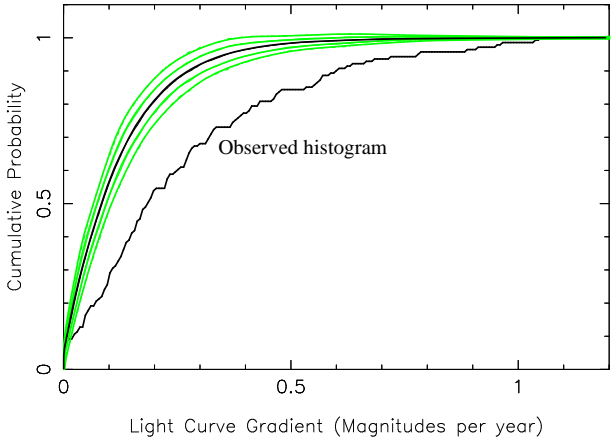}
\caption{\label{noise}The cumulative histogram of the monitoring data (labeled)together with the cumulative histogram for simulations where the only fluctuation is due to that from the assigned errors (dark line). Also shown are the $\pm1\sigma$ and $\pm2\sigma$ levels (light lines). In this case the assigned errors were described by a Gaussian with a standard deviation of the quoted observational random error.}
\end{figure}

The first thing to note about this interpretation of the monitoring data is that there is a significant signal of uncorrelated fluctuations. The level of noise in the sample can be determined by calculating the fluctuations that result from the application of the observational uncertainty and sampling rate to a flat light curve. Figure \ref{noise} shows the cumulative histogram of the rates of change of the difference in the image magnitudes of Q2237+0305 together with cumulative histograms for the mean, $\pm1\sigma$ and $\pm 2\sigma$ values for the cumulative histogram of derivatives of fluctuations due to observational error only. The spread in this distribution results from the short sample time scale of the data for Q2237+0305. The uncorrelated fluctuation in the low gradient regime is very strong evidence for microlensing, particularly when added to that from the large independent fluctuations or microlensing events that have been observed (Irwin et al. 1989, Corrigan et al. 1991).   

\subsection{Microlensing Models for Q2237+0305}

\begin{figure}
\vspace*{80mm}
\includegraphics{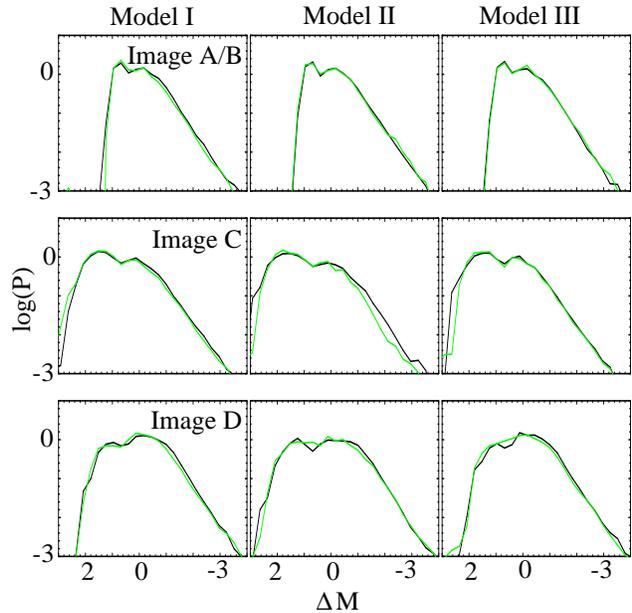}
\caption{\label{ampdist} The magnification distributions for our simulations of the images in Q2237+0305. The dark and light lines represent simulations that have positive ($\gamma>0$) and negative ($\gamma<0$) shear respectively.}
\end{figure}

\begin{table*}
\begin{center}
\caption{\label{tab_field} The values of average magnification, and the number of stars used in each of the models. Values are shown for each set of image parameters and microlens mass model. In the case of the number of stars used, the first value corresponds to the simulations with positive shear and the second value to simulations with negative shear. The theoretical magnifications for images A/B, C and D are $\mu_{th}$ = 4.00, 2.45 and 4.90 respectively.}
\begin{tabular}{|c|c|c|c|c|c|c|c|c|c|}

\hline
      & \multicolumn{3}{c}{Image A/B} & \multicolumn{3}{c}{Image C} & \multicolumn{3}{c}{Image D} \\
Model & Model $\mu_{av}$ & \multicolumn{2}{c}{$N_{*}$}  & Model $\mu_{av}$ & \multicolumn{2}{c}{$N_{*}$} & Model $\mu_{av}$ & \multicolumn{2}{c}{$N_{*}$} \\ \hline\hline
$I$   & 4.02$\pm$0.15    & 2926 & 500 & 2.48$\pm$0.11    & 2782 & 735 & 4.80$\pm$0.11    & 3178 & 593 \\
$II$  & 3.97$\pm$0.11    & 4104 & 1549 & 2.31$\pm$0.20    & 4420 & 2403 & 4.71$\pm$0.40    & 13033 & 5656 \\
$III$ & 3.95$\pm$0.14    & 1105 & 500 & 2.38$\pm$0.16    & 1164 & 597 & 4.85$\pm$0.24    & 3463 & 1383 \\\hline \\\\
  
\end{tabular}
\end{center}
\end{table*}

 To model microlensing in Q2237+0305 we adopt the lensing galaxy model of Schmidt, Webster \& Lewis (1998). This model includes the effect of the bar and produces the following microlensing parameters: image A ($\kappa=0.36,\,|\vec{\gamma}|=0.40$), image B ($\kappa=0.36,\,|\vec{\gamma}|=0.40$), image C ($\kappa=0.69,\,|\vec{\gamma}|=0.71$) and image D ($\kappa=0.59,\,|\vec{\gamma}|=0.61$). We consider four models for the distribution of mass in our microlensing model. These models are as follows:

\noindent Model $I$: The microlenses are distributed in the range $0.1M_{\odot}<m<1.0M_{\odot}$ with a Salpeter mass function described by $p(m)dm\propto m^{-2.35}dm$.

\noindent Model $II$: The microlenses are distributed in the range $0.1M_{\odot}<m<10.0M_{\odot}$ with a Salpeter mass function described by $p(m)dm\propto m^{-2.35}dm$.

\noindent Model $III$: The microlenses are each assigned a mass of $m=1.0M_{\odot}$.

\noindent There is no continuously distributed matter in any of our models. None of our models contain low mass ($m<0.1M_{\odot}$) compact objects. This choice follows the results of Alcock et al. (1997) who claim from MACHO experiments that the mean mass of dark compact objects in the Milky Way halo is between $\sim0.2$M$_{\odot}$ and $\sim0.8$M$_{\odot}$. In addition, Schmidt \& Wambsganss (1998) place limits on the composition of the halo in Q0957+561 from the observed lack of microlensing. Where the halo is made entirely of compact objects they rule out a typical MACHO mass of $< 0.001$M$_{\odot}$. We note here that (as discussed below) a smaller measure of transverse velocity will result from a model that contains smaller compact objects. Our upper limits will therefore be unaffected by this choice. 
 
 Our microlensing models assume that the mass function and range in the bulge is the same at the position of each of the four images. The orientation of the bar lies approximately along the A-B image axis (Yee 1988). We note therefore that this assumption may be rendered invalid if the bar population varies from that in the rest of the bulge. The images are approximately orthogonal with respect to the galactic centre, and so the shear polar in images A and B has a direction which is approximately orthogonal to that in images C and D. For each microlensing model we consider two cases, (1) the shear in images A and B is parallel with the transverse velocity (described by $\gamma>0$), and (2) in images C and D is parallel with the transverse velocity (described by $\gamma<0$). These bracket the range of possibilities for the orientation between the source trajectory and the galaxy.

For each of the four mass models, and for both positive ($\gamma>0$) and negative ($\gamma<0$) applied shear, 30 light curves were computed for the microlensing parameters corresponding to each image. Each individual light curve had a length of 10 Einstein radii. Table \ref{tab_field} shows the numbers of point masses required in these models to collect the prescribed macro-image flux. The values on the left and right refer to simulations that have positive shear ($\gamma>0$) and negative shear ($\gamma<0$) respectively.
 Table \ref{tab_field} also shows the mean magnification found from each of the models together with the theoretical average. These values  demonstrate that our models are accounting for an appropriate percentage of the macro image flux. The set of simulations for each sign of shear were divided up into 3 subsets (of length 100 Einstein radii each).  The quoted errors were calculated from a $\sigma_{n-1}$ standard deviation between the resulting 6 values.
For comparison with previous publications (eg. Lewis \& Irwin 1995) figure \ref{ampdist} shows the magnification distributions obtained from our models. As required we find that the distributions are independent of the sign of the shear.

 Models of the observed light curves were computed using a sampling rate identical to that of the set of observations. This sampling rate is applied to the various model light curves in combination with the simulated observational errors. For each of the models considered, an ensemble of 1500 simulations was produced for statistical comparison with the observed light curve of Q2237+0305 through the method described in section \ref{ks}. 

\begin{figure*}
\vspace*{200mm}
\includegraphics{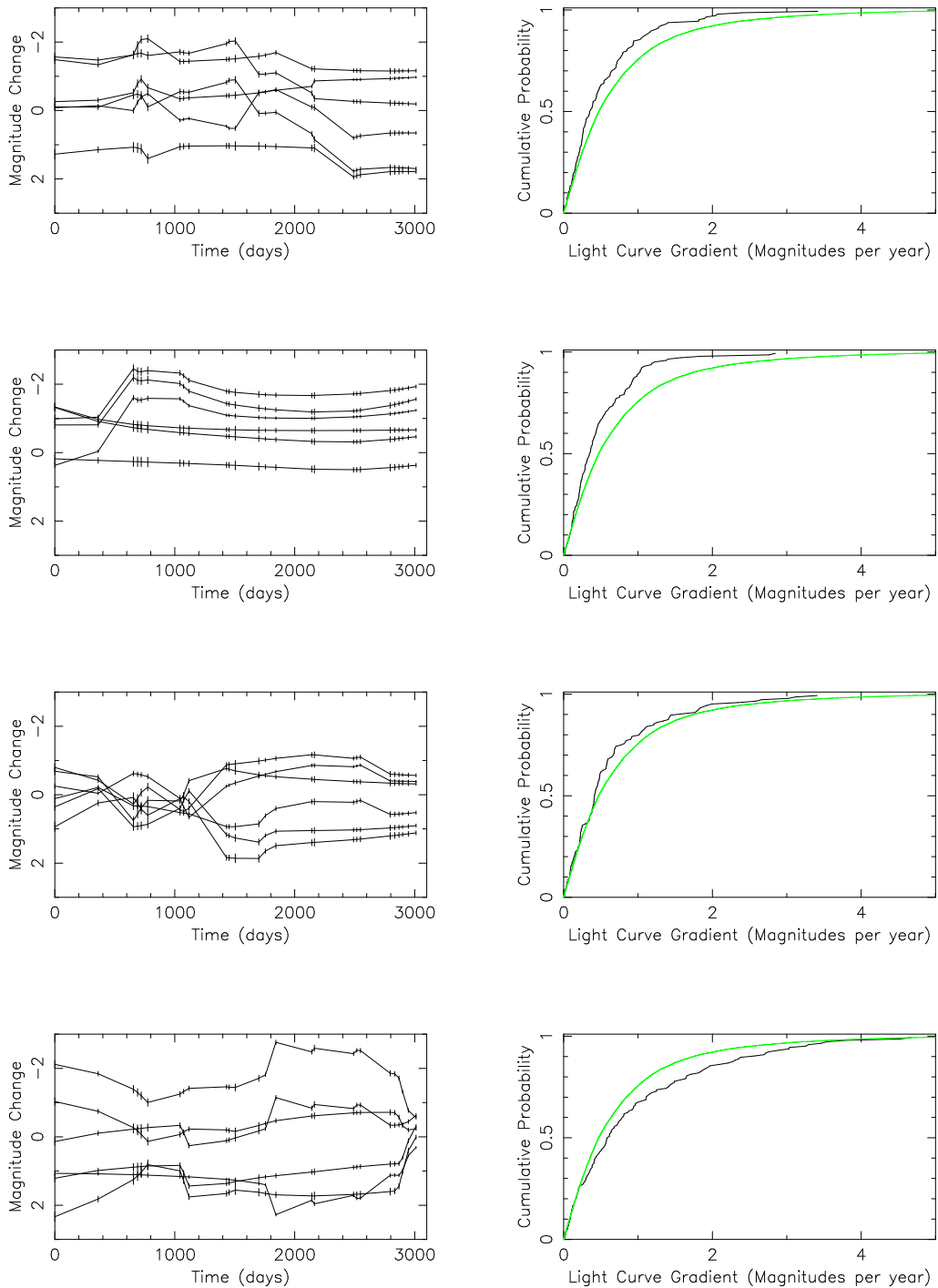}
\caption{\label{simul_hist}Three typical simulations for a point source with a type $I$ model. A simulated observational random error of $1\sigma$ for images A and B, and $\frac{1}{2}\sigma$ for images C and D is plotted at each point. Left: The six differences between model light curves of the images of Q2237+0305. Right: The resulting cumulative histogram (dark line) together with the time average histogram (light line). In this case the transverse velocity was $400\,km\,sec^{-1}$. The optical depth and shear values were those of Q2237+0305, with $\gamma_{A}, \gamma_{B}>0$ and $\gamma_{C}, \gamma_{D}<0$.}
\end{figure*}

Figure \ref{simul_hist} shows four examples of modelled variations in the 6 image magnitude differences, together with the resulting cumulative histograms. The figure also shows the time averaged mean calculated from the entire sample (light line). In this case the effective transverse velocity in the models was 400$\,km\,sec^{-1}$. Figure \ref{simul_hist} demonstrates that such a short monitoring period can mean that the number of large amplitude fluctuations varies wildly depending on the period of light curve sampled.

\section{Results}
\label{results}
\subsection{Effective Transverse Velocity}

\begin{table*}
\begin{center}
\caption{\label{tab_1} The effective transverse velocities ($km\,sec^{-1}$). The first and second numbers in each column refer to models where $\gamma_{A},\gamma_{B}>0$, $\gamma_{C},\gamma_{D}<0$ and $\gamma_{A},\gamma_{B}<0$, $\gamma_{C},\gamma_{D}>0$ respectively.}
\begin{tabular}{|c|c|c|c|c|c|c|c|c|c|c|c|c|c|c|c|c|c|c|c|}
\hline
Model  & \multicolumn{2}{c}{Error $(\pm .02)$}    &\multicolumn{4}{c}{$V_{lower}(95\%)$}& \multicolumn{4}{c}{$V_{mean}$}    &\multicolumn{4}{c}{$V_{upper}(95\%)$}  & \multicolumn{4}{c}{$V_{upper}(99\%)$} \\
Type    & $\sigma_{A,B}$  & $\sigma_{C,D}$  & \multicolumn{2}{c}{point} & \multicolumn{2}{c}{0.1$\eta_{o}$} & \multicolumn{2}{c}{point} & \multicolumn{2}{c}{0.1$\eta_{o}$} &           \multicolumn{2}{c}{point} & \multicolumn{2}{c}{0.1$\eta_{o}$} &       \multicolumn{2}{c}{point} & \multicolumn{2}{c}{0.1$\eta_{o}$}  \\ \hline\hline
$I$ & 1 & $\frac{1}{2}$       & 10 & 20   &   20 & 20         & 70 & 80   &  70 & 70          & 340 & 330  &   310 & 290        & 470 & 470  &  450 & 420         \\
$I$ & 2 & 1       &  50 & 50  &    20 & 20        & 200 & 190  &  160 & 150         & 520 & 470  &   460 & 400        & 740 & 640  &  680 & 590          \\
$II$ & 1 & $\frac{1}{2}$    & 20 & 20   &   20 & 20         & 120 & 100  &   90 & 90         & 470 & 400  &  440 & 370         & 740 & 550  &  630 & 500          \\
$II$ & 2 & 1     &  70 & 50  &    40 & 40        & 280 & 250  &  220 & 210         & 710 & 600  &  650 & 520         & 990 & 810  &   930 & 760         \\
$III$ & 1 & $\frac{1}{2}$ & 20 & 20   &   20 & 20         & 200 & 160  &  170 & 120         & 650 & 590  &  620 & 540         & 930 & 890    &  890 & 810          \\
$III$ &  2 & 1     & 110 & 100  &    70 & 70        & 410 & 380 &  370 & 280         & 970 & 880  &   870 & 790        & 1330 & 1150   &    1250 & 1030       \\ \hline
\end{tabular}
\end{center}
\end{table*}

\begin{figure}
\vspace*{65mm}
\includegraphics{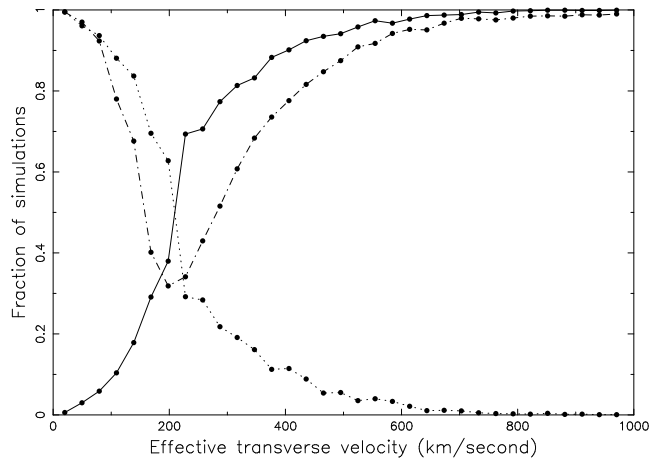}
\caption{\label{limits}Plots of the probability of finding a model - average model KS difference greater than the observation - average model KS difference as a function of effective transverse velocity. The solid line and dotted lines correspond to the upper and lower limits respectively. The dot-dashed line corresponds to the mean. In this case the microlensing model was of type $I$. The optical depth and shear values were those of Q2237+0305, with $\gamma_{A}, \gamma_{B}>0$ and $\gamma_{C}, \gamma_{D}<0$. The observational random error was taken to be $2\sigma$ for images A and B, and $1\sigma$ for images C and D}
\end{figure}

 Figure \ref{limits} shows plots of effective transverse velocity vs. percentage of simulations that are more consistent with the model time average than the observations in terms of the upper limit, lower limit, and mean KS statistics (solid, dotted, and dot-dashed lines) than the observations. This plot measures the most likely as well as the upper and lower limits to the effective transverse velocity in the case of a type $I$ microlensing model. Similar plots were produced in each of the models considered. In each of the three models, the effective transverse velocities were found for which the cumulative histogram of observations was less consistent with the time averaged model distribution than $95\%$ and $99\%$ of the ensemble of model simulations ($V_{upper}(95\%)$ and $V_{upper}(99\%)$) in the case of the upper limit, and 95\% of the ensemble simulations in the case of the lower limit ($V_{lower}(95\%)$). The effective transverse velocity at which the histogram of observations was more consistent with the time average than the greatest number of models was also found ($V_{mean}$). These transverse velocities were found in the cases of point, $0.02\eta_{0}\:(2.6\times 10^{15}cm)$ and $0.10\eta_{0}\:(1.3\times 10^{16}cm)$ diameter sources. The calculations of the finite source light curves were made with the one dimensional approximation described in section \ref{source_sixe}. Each of the above measurements were made using the both assumptions of simulated observational error discussed previously.

\begin{figure}
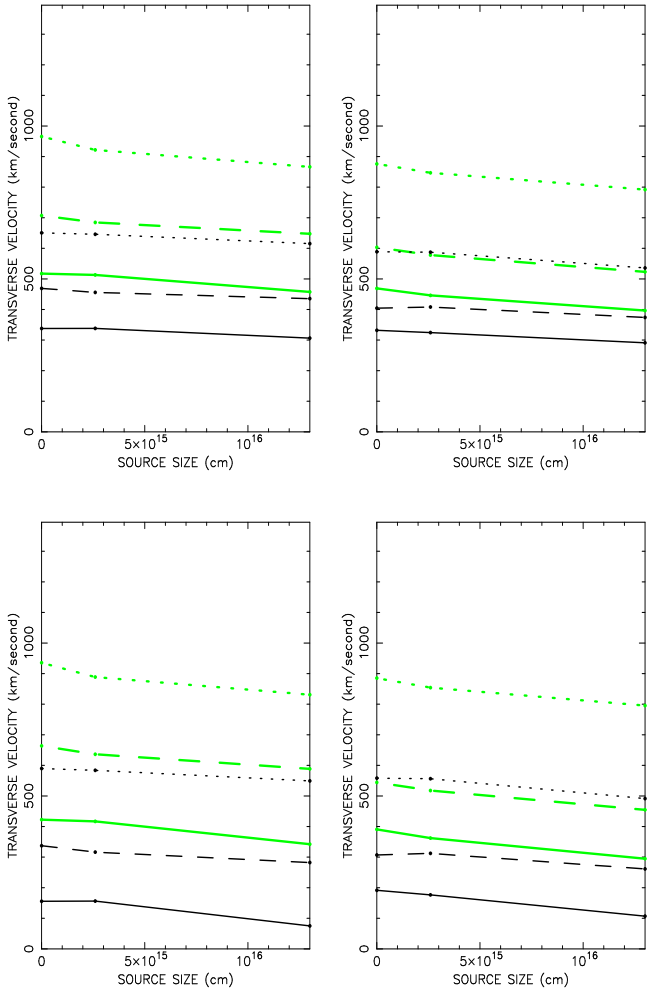

\vspace*{135mm}
\includegraphics{figures/fig12a.ps}
\includegraphics{figures/fig12b.ps}
\caption{\label{upper_lim_90}Top: Plots of the 95\% upper limit to the effective transverse velocity, and Bottom:  Plots of the 95\% upper limit to the physical transverse velocity as a function of source size for model $I$ (solid line), model $II$ (dashed line) and model $III$ (dotted line). The light lines are calculated using a simulated observational error of $2\sigma$ for images A and B, and $1\sigma$ for images C and D, while the dark lines correspond to errors twice as large. The left hand panels correspond to the case where $\gamma_{A},\gamma_{B}>0$,and  $\gamma_{C},\gamma_{D}<0$}
\end{figure}

 Table \ref{tab_1} displays the results obtained for the effective transverse velocity ($V$) computed using the point and $0.1\eta_{o}$ diameter sources. The results for $V_{upper}(95\%)$ at all three sizes considered are presented in the top panel of Figure \ref{upper_lim_90}. The left and right hand plots represent the cases of $\gamma_{A},\gamma_{B}>0$ and  $\gamma_{A},\gamma_{B}<0$. In these diagrams the solid, dashed and dotted lines represent models $I$, $II,$ and $III$ respectively. The light lines correspond to the quoted observational errors being treated as 2$\sigma$ in images A and B, and 1$\sigma$ in images C and D, while the dark lines correspond to errors of 1$\sigma$ in images A and B, and $\frac{1}{2}\sigma$ in images C and D. From these plots we see that the models yield estimates of the effective transverse velocity that are qualitatively consistent. It is apparent from the plot however that it is the model of microlenses chosen that provides the largest uncertainty. The ranges of error considered have altered the estimated limits on effective transverse velocity by $\sim30\%$, while the source size and the direction of the transverse velocity provide relatively unimportant contributions to the systematic uncertainty in the limits obtained.

Models with a smaller mean mass, and therefore a larger number density of caustics measure a lower effective transverse velocity. The Einstein radius of a point mass in normalised units is $\sqrt{m}$. The effective transverse velocities measured using two models $a$ and $b$ that have identical macro parameters ($\kappa,\gamma$), and mass limits whose ratios are equal are related by a factor of $\sqrt{\langle m_{a}\rangle /\langle m_{b}\rangle}$. It is therefore interesting to note the correspondence between this ratio in the models we have considered and the effective transverse velocities measured. 

\vspace{2mm}

\noindent In the case of models $I$ and $II$:

\noindent$\sqrt{\langle m_{II}\rangle / \langle m_{I}\rangle }=1.2$, $V_{mean}(II)/V_{mean}(I)\sim1.3$.

\vspace{2mm}

\noindent In the case of models $I$ and $III$:

\noindent$\sqrt{\langle m_{III}\rangle / \langle m_{I}\rangle}=2.1$, $V_{mean}(III)/V_{mean}(I)\sim2$. 

\vspace{2mm}

\noindent The good fit of these values demonstrates that the measurement of effective transverse velocity is a sensitive function only of the mean microlens mass. This result is consistent with Witt, Kaiser \& Refsdal (1993) and with Lewis \& Irwin (1996) who found that the characteristic time-scale for variability scales as $\sqrt{\langle m\rangle}$. In addition, figure \ref{ampdist} demonstrates that the magnification distribution is independent of the mass function and mean microlens mass (Lewis \& Irwin 1995). For these reasons we have not explored models with mass functions described by a larger range of power laws or mass limits.

 A major unknown in the problem is the direction of the source trajectory. Simulations by Witt, Kayser \& Refsdal (1993) and others show that for image A the rate of HMEs is significantly higher for a source moving parallel to the shear ($\gamma>0$), similar results are obtained for images C and D, although the effect is more pronounced. A combination of this effect in each of the 4 images of Q2237+0305 means that the shape of the derivative histogram from the difference light curves is dependent on the direction of the transverse velocity. Figure \ref{upper_lim_90} shows that the measurement of effective transverse velocity is slightly smaller where the sign of the shear in images $A$ and $B$ is negative.

While the results are model dependent, some general conclusions are readily apparent. Firstly, in all models considered the upper limits of effective transverse velocity are of the same order as, and the most likely values significantly lower than the previously assumed value for the transverse velocity of 600$\,km\,sec^{-1}$. This suggests that the transverse velocity and stellar velocity dispersion have similar magnitudes, and that microlensing by random motion of stars may play a significant role in producing the observed continuum variation.  
In addition, the lower limit to the effective transverse velocity ($V_{lower}(95\%)$) is greater than zero for all models considered. This result is required as evidence that there is microlensed continuum variability which is above the level generated by observational noise, and is a quantification of figure \ref{noise}.

\subsection{The Transverse Velocity}
\begin{table*}
\begin{center}
\caption{\label{tab_2}The transverse velocities ($km\,sec^{-1}$). The first and second numbers in each column refer to models where $\gamma_{A},\gamma_{B}>0$, $\gamma_{C},\gamma_{D}<0$ and $\gamma_{A},\gamma_{B}<0$, $\gamma_{C},\gamma_{D}>0$ respectively.}
\begin{tabular}{|c|c|c|c|c|c|c|c|c|c|c|c|c|c|c|c|c|c|c|}
\hline
Model   & \multicolumn{2}{c}{Error $(\pm .02)$}   &  \multicolumn{4}{c}{$v_{lower}(95\%)$} &  \multicolumn{4}{c}{$v_{mean}$}    & \multicolumn{4}{c}{$v_{upper}(95\%)$} & \multicolumn{4}{c}{$v_{upper}(99\%)$}                \\
Type    & $\sigma_{A,B}$   & $\sigma_{C,D}$   & \multicolumn{2}{c}{point} & \multicolumn{2}{c}{0.1$\eta_{o}$} & \multicolumn{2}{c}{point} & \multicolumn{2}{c}{0.1$\eta_{o}$} & \multicolumn{2}{c}{point} & \multicolumn{2}{c}{0.1$\eta_{o}$} & \multicolumn{2}{c}{point} & \multicolumn{2}{c}{0.1$\eta_{o}$}  \\ \hline\hline
$I$ & 1 & $\frac{1}{2}$       &     0 & 0      & 0 & 0  &    0 & 0        &    0 & 0       & 160& 190  &    80& 110        & 360& 390  &  340& 330          \\
$I$ & 2 & 1      &  0 & 0 &   0    &     0   & 0  &    0        &    0   &  0    &  420 & 390  &   340 & 300        & 670 & 580  &  610 & 540          \\
$II$ & 1 & $\frac{1}{2}$     &    0&   0      & 0 & 0  &    0 & 0        &   0 & 0        & 340& 310  &   280& 260        &  700& 480 &  570& 440          \\
$II$ & 2 & 1    &   0&  0&    0    &  0      & 0   &   0        &   0     &   0  &   660 & 540  &  590 & 450         &  980 & 760 &  920 & 710          \\
$III$ & 1 & $\frac{1}{2}$  &     0 & 0      & 0 & 0  &  0 & 0          &  0&  0         & 590& 560  &   550& 490        & 900& 890    &  860& 820          \\
$III$  & 2 &  1    &   0&  0&    0    &  0      &  310 & 280 &   250 & 110        & 940 & 890  &   830 & 800        &    1320 & 1180 &    1230 & 1050       \\ \hline

\end{tabular}
\end{center}
\end{table*}

 The conversion from an effective transverse velocity to a physical one was made using the upper limit of relationships such as the one shown in figure \ref{vel_convert}. This corresponds to the subtraction of a lower limit to the contribution of proper motions to the microlensing rate. The subtraction of the proper motion component is of a similar magnitude for all models. The range of velocities is however increased because the difference between effective and true transverse velocity is less for larger transverse velocities (see figure \ref{vel_convert}).The results are presented in Table \ref{tab_2}. The measured values of transverse velocity are higher for smaller source sizes and smaller simulated errors. Importantly, the most likely transverse velocity is significantly lower than the expected value of $600\,km\,sec^{-1}$, with values that are zero in most cases considered. This zero value does not signify a lack of microlensing, but rather that there is not an excess of microlensed fluctuation above that which is expected from proper motions alone.

 The lower plots in figure \ref{upper_lim_90} show the values of transverse velocity corresponding to the effective transverse velocity plotted in the upper part of the figure. From the results of these models we estimate a maximum expected value for the galactic transverse velocity. From the results of the 95\% upper limit we place the expected galactic transverse velocity at $v_{t}<500\,km\,sec^{-1}$. This upper limit demonstrates that the galactic transverse velocity is probably significantly lower than the value that has been assumed for analyses of microlensing in Q2237+0305 in the past.

\section{Conclusion}

Through consideration of the distribution of light curve derivatives we find that the ten years of existing monitoring data for Q2237+0305 contains a statistically significant level of uncorrelated and therefore microlensed variation in the continuum flux magnitudes of its four images. This microlensing
is predominantly at a level below that which would be considered to be a HME. Through consideration of the total rate of microlensed variation, and the contribution to microlensing of stellar proper motion we have placed an upper limit on the galactic transverse velocity of $v_{t}< 500\,km\,sec^{-1}$. This measurement is not significantly dependent on the size or intensity profile of the continuum region, or on the direction of the source trajectory. The choice of microlensing model in combination with the reliability of the error estimates produces the greatest uncertainty in the problem, however the estimate is qualitatively consistent over a range of possible models. This data suggests that the transverse velocity of the lensing galaxy of Q2237+0305 is likely to be low, in which case the observed microlensing is primarily due to the proper motion of microlenses in the bulge. 

 Analyses such as the one discussed here will provide more rigorous limits when data from a longer monitoring period becomes available. The statistical significance of a measurement of transverse velocity will not be substantially aided by a sampling rate smaller than about 1 point per month. However, knowledge of the errors involved is important to obtaining a reliable limit.

\label{lastpage}


\begin{thebibliography}{}

\bibitem[\protect\citename{Alcock et al. }1997]{AL97}
Alcock, C., Allsman, R. A., Alves D., et al. (The MACHO Collaboration), 1997, Ap. J., 486, 697
 
\bibitem[\protect\citename{Corrigan et al. }1991]{CO91}
Corrigan et al., 1991, Astron. J., 102, 34
 
\bibitem[\protect\citename{Foltz et al.  }1992]{FO92}
Foltz, C. B., Hewitt, P. C., Webster, R. L., Lewis, G. F., 1992, Ap. J., 386, L43
 
\bibitem[\protect\citename{Irwin et al.  }1989]{IR89}
Irwin, M. J., Webster, R. L., Hewitt, P. C., Corrigan, R. T., Jedrzejewski, R. I., 1989, Astron. J.
, 98, 1989
 
\bibitem[\protect\citename{Katz, Balbus \& Paczynski }1986]{KA86}
Katz, N., Balbus, S., Paczynski, B., 1986, Ap. J., 306, 2
 
\bibitem[\protect\citename{Kent \& Falco }1988]{KE88}
Kent, S. M., Falco, E. E., 1988, Astron. J., 96, 1570

\bibitem[\protect\citename{Kundic \& Wambsganss }1993]{KU93}
Kundic, T., Wambsganss, J., 1993, Ap. J., 404, 455
 
\bibitem[\protect\citename{Kundic, Witt \& Chang }1993]{KU93b}
Kundic, T., Witt, H. J., Chang, K., 1993, Ap. J., 409, 537
 
\bibitem[\protect\citename{Lewis \& Irwin} 1995]{LE95}
Lewis, G. F., Irwin, M. J., 1995, MNRAS 276, 103

\bibitem[\protect\citename{Lewis \& Irwin} 1996]{LE96}
Lewis, G. F., Irwin, M. J., 1996, MNRAS 283, 225

\bibitem[\protect\citename{Lewis et al.   }1993]{LE93}
Lewis, G. F., Miralda-Escude, J., Richardson, D. C., Wambsganss, J., 1993, MNRAS, 261, 647
 
\bibitem[\protect\citename{Mould et al.} 1993]{MO93}
Mould, J. R., Akeson, R. L., Bothun, G. D., Han, M., Huchra, J. P., Roth, J., Schommer, R. A., 1993, Ap. J., 409, 14
 
\bibitem[\protect\citename{$\O$stensen et al.}1996]{OS96}
$\O$stensen, R. et al. 1996, Astron. Astrophys., 309, 59

\bibitem[\protect\citename{Rauch \& Blandford } 1991]{RA92}
Rauch, K. P., Blandford, R. D., 1991, Ap. J. 1991, 381, L39

\bibitem[\protect\citename{Rees } 1984]{R84}
Rees, M., J., 1984, ARA\&A, 21, 471

\bibitem[\protect\citename{Schmidt \& Wambsganss}1998]{SC98}
Schmidt, R. W., Wambsganss, J., 1998, Astron. Astrophys., 335, 379

\bibitem[\protect\citename{Schmidt, Webster \& Lewis }1998]{SC98a}
Schmidt, R. W., Webster, R. L., Lewis, G. F. 1998, MNRAS, 295, 488
 
\bibitem[\protect\citename{Schneider et al. }1988]{SC88}
Schneider, D. P., Turner, E. L., Gunn, J. E., Hewitt, J. N., Schmidt, M., Lawrence, C. R., 1988, Astron. J., 95, 1619 

\bibitem[\protect\citename{Schramm et al.}1993]{SC93}
Schramm, T., Kayser, R., Chang, K, Nieser, L., Refsdal, S
., 1993, Astron. Astrophys., 268, 350

\bibitem[\protect\citename{Wambsganss, Paczynski \& Katz }1989]{WA89}
Wambsganss, J., Paczynski, B., Katz, N., 1989, Ap. J., 352, 407 
 
\bibitem[\protect\citename{Wambsganss, Paczynski \& Schneider }1990]{WA90}
Wambsganss, J., Paczynski, B., Schneider, P., 1990, Ap. J., 358, L33

\bibitem[\protect\citename{Wambsganss \& Kundic } 1995]{WA95}
Wambsganss, J., Kundic, T., 1995, Ap. J., 450, 19
 
\bibitem[\protect\citename{Witt, Kayser \& Refsdal }1993]{WI93}
Witt, H. J., Kayser, R., Refsdal, S. 1993, Astron. Astrophys.,268, 501
 
\bibitem[\protect\citename{Witt}1993]{WI93a}
Witt, H. J., 1993, Ap. J., 430, 530

\bibitem[\protect\citename{Witt \& Mao }1994]{WI94}
Witt, H. J., Mao, S., 1994, Astron. J., 429, 66

\bibitem[\protect\citename{Wyithe \& Webster }1999]{WY99a}
Wyithe, J. S. B, Webster, R. L., 1999, MNRAS accepted
 
\bibitem[\protect\citename{Wyithe, Webster \& Turner }1999]{WY99b}
Wyithe, J. S. B, Webster, R. L., Turner, E. L., 1999, MNRAS submitted.

\bibitem[\protect\citename{Yee }1988]{YE88}
Yee, H. K. C., 1988, Ap. J., 95, 1331

\end{thebibliography}
\end{document}